\documentclass[11pt]{article}
\usepackage{moriond,epsfig}

\def\be{\begin{equation}}
\def\ee{\end{equation}}
\def\bea{\begin{eqnarray}}
\def\eea{\end{eqnarray}}

\def\gluino{\ensuremath{\tilde{g}}}
\def\squark{\ensuremath{\tilde{q}}}

\def\sbottom{\ensuremath{\tilde{b}}}

\newcommand{\anactualptmissvecwithapinit}{{\vec{P}}_\mathrm{T}^\mathrm{miss}}

\newcommand{\ourvecptmiss}{\ensuremath{\anactualptmissvecwithapinit}}
\newcommand{\ourmagptmiss}{\MET}

\newcommand{\mttwo}{\ensuremath{m_\mathrm{T2}}}
\newcommand{\ourdeltaphishort}{\Delta\phi}
\newcommand{\ourdeltaphifull}{\ourdeltaphishort(\textrm{jet},\ourvecptmiss)_\mathrm{min}}

\def\MET{\ensuremath{E_{\mathrm{T}}^{\mathrm{miss}}}} 
\def\met{\ensuremath{E_{\mathrm{T}}^{\mathrm{miss}}}} 

\newcommand{\meff}{\ensuremath{m_{\mathrm{eff}}}}

\begin{document}
\vspace*{4cm}
\title{SUSY searches at ATLAS}

\author{Sascha Caron for the ATLAS collaboration}

\address{Physikalisches Institut, University of Freiburg
Hermann-Herder Str.3 , Freiburg, Germany}

\maketitle\abstracts{
First ATLAS searches for signals of Supersymmetry in proton-proton collisions at the LHC are presented. These searches are performed in various
channels containing different lepton and jet multiplicities in the final states; the full data sample recorded in the 2010 LHC run, corresponding to an integrated luminosity of $35$ $\rm{pb}^{-1}$, has been analysed. The limits on squarks and gluinos are the most stringent to date.}

\section{Introduction}
Supersymmetry (SUSY) is one of the most favoured candidates for new physics, predicting a new symmetry between fermions and bosons and therefore a large number of new particles~\cite{Martin:1997ns}. There are various reasons why these new SUSY particles are expected to be not too heavy, but accessible with TeV scale energies. SUSY would reduce the so-called {\it fine-tuning problem} of the Standard Model (SM) and it
indicates that the three forces of the Standard Model are unified at very high energies.
On the other hand SUSY particles have not been discovered so far and thus need to be heavier than their SM counterparts. 
In addition a new multiplicative quantum number called R-parity is introduced to forbid strong lepton and baryon number violating terms in the SUSY Lagrangian leading to too rapid proton decay. 
If the R-parity quantum number is conserved and SUSY particles are heavier than their SM partners, then SUSY predicts a large amount of Dark Matter in the universe. These reasons make the search
for SUSY particles a major and important part of the LHC physics program. 
New SUSY particles could be discovered or if nothing is found at the LHC the SUSY solution to the SM shortcomings will become very unlikely.

At the LHC new SUSY particles are produced in pairs (if R-parity is conserved) and each decays usually via several intermediate steps (cascade decay) to the lightest SUSY particle (LSP). 
The LSP is only weakly interacting due to cosmological arguments and leads to the most characteristic feature of these SUSY events, which is missing transverse momentum. Since the SUSY breaking mechanism is unknown the mass pattern of the SUSY particles cannot be predicted. 
The search strategy needs therefore to be quite generic or SUSY model parameter independent.
A generic search strategy for R-parity conserving 
SUSY signatures would include the selection of events with large missing transverse energy and reconstructed particles with large transverse momentum. 
At the LHC these objects are predominantly jets since the coupling strength of the strong force would cause an abundance of squarks and gluinos if these particles are not too heavy. 
Squarks or gluinos will cascade decay to jets, several leptons or photons depending on the SUSY parameters and missing transverse momentum caused by the LSPs.
The searches for SUSY signatures with R-parity conservation are performed by searching for more events than expected in a number of different channels. These channels explore a large variety of possible signals, e.g. ATLAS studies various different jet (2,3,4) and lepton (0,1,2,3) multiplicities.
The main challenge in these searches (and most of the work) is to reliably control the Standard Model background expectations. 
In the following
we assume mostly the 5-parameter mSUGRA as a ``general'' model for R-parity conserving SUSY. 
Model-independent limits on an 
{\it effective} cross section for new processes in the
signal region, including the effects of
experimental acceptance and efficiency have also been derived. 
They can be used to exclude any model of new physics leading to a larger {\it effective} 
cross section.

All data presented in this summary are 
taken in the year 2010 at the ATLAS experiment~\cite{Aad:2008zzm} 
in LHC proton-proton collisions at 7 TeV centre-of-mass energy,
using a data-set corresponding to an integrated luminosity of about $35 \rm{pb}^{-1}$. 

\section{Searches with Jets, a Lepton and Missing Transverse Momentum (1 lepton channel)}

\begin{figure}
\epsfig{figure=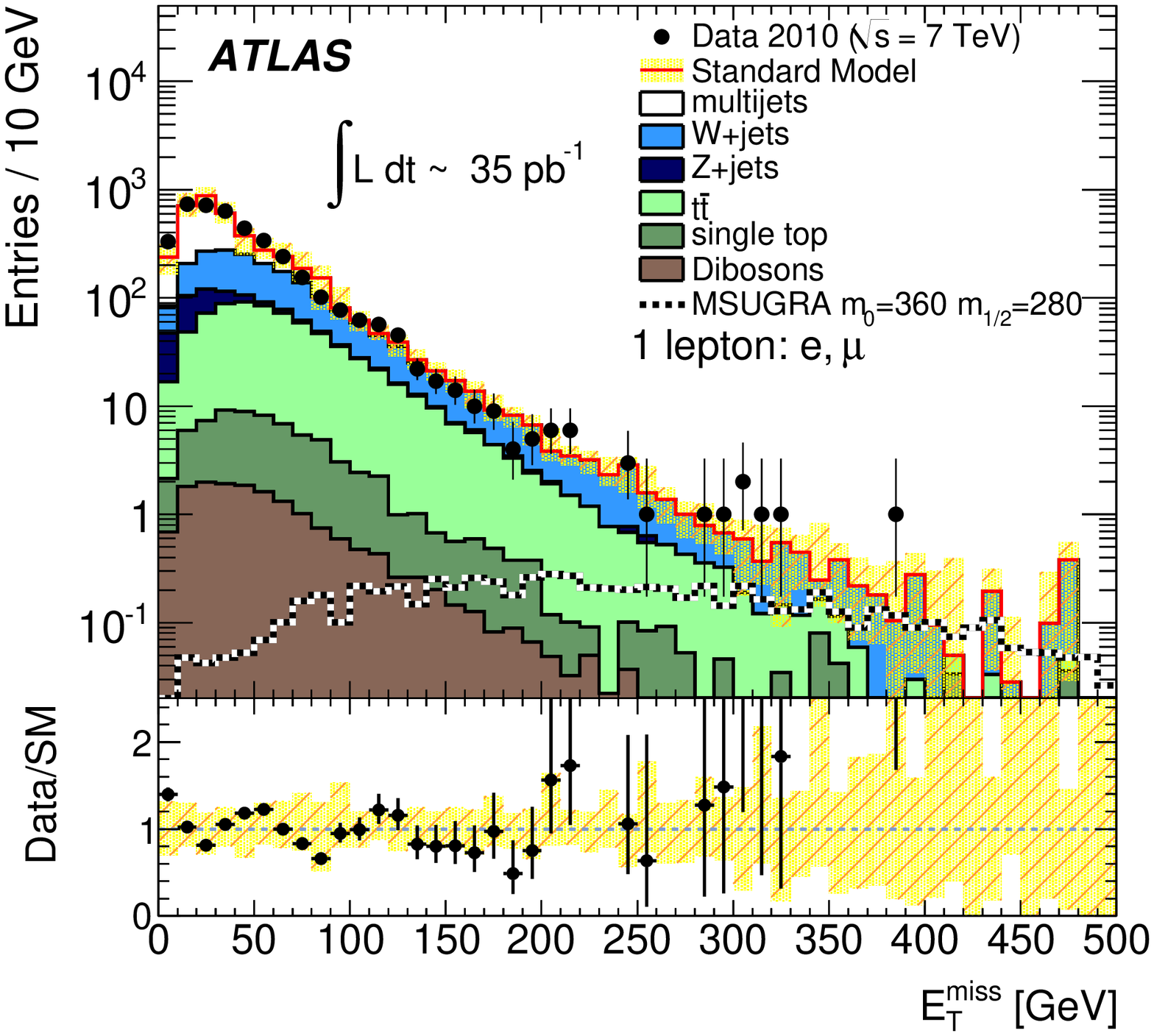,width=7.5cm}
\epsfig{figure=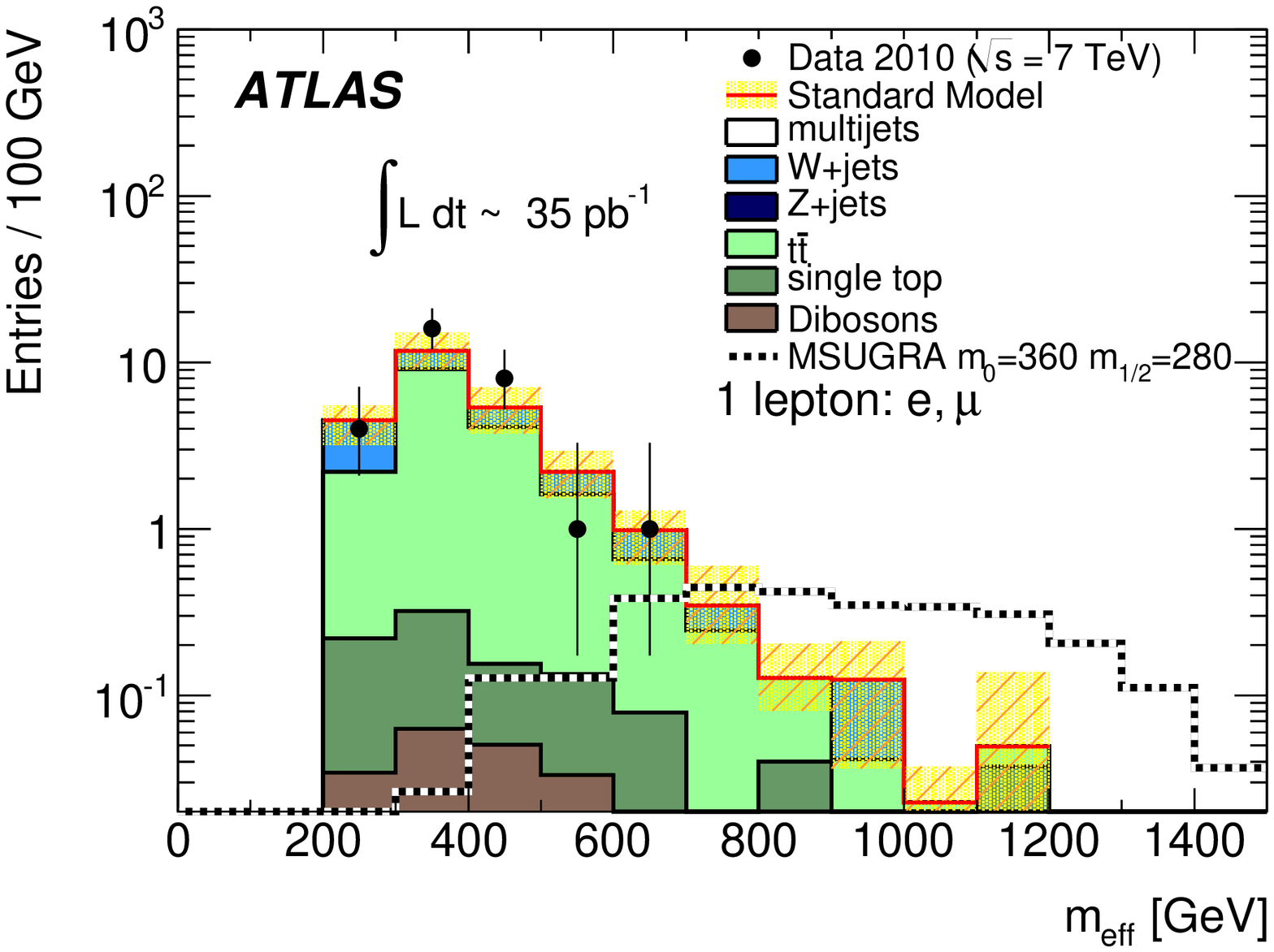,width=8.5cm}
\caption{$E_T^{miss}$ distribution after lepton and jet selection (left figure). 
Effective mass distribution after final selection criteria except for the cut on the effective mass itself (right figure). The 
plots are made for the electron and muon channel combined.
Yellow bands indicate the uncertainty on the Monte Carlo prediction 
from finite Monte Carlo statistics and from the jet energy scale uncertainty.} 
\label{fig:1lepton}
\end{figure}

The search in a channel with jets, exactly one muon or electron and significant
missing transverse momentum \met constitutes the first ATLAS SUSY result
~\cite{Aad:2011hh}.
This channel could be studied first due to the large reduction of
the potentially dangerous QCD multijet background. Requiring one muon or electron
reduces this background by several orders of magnitude.
The left plot of Figure 
\ref{fig:1lepton} shows the missing transverse momentum distributions after requiring
three jets and one muon or electron. The data is compared to the prediction from Monte Carlo.
Only the QCD prediction from PYTHIA was scaled with a k-factor depending on the muon or electron selection. The figure shows that at high $\met$ the main background comes from W+jets events
and events from top pair production. The data is in agreement with the expectations.
The signal region is defined by cuts on  $\met >125$~GeV, $\met >0.25 \meff$
and $\meff >500$~GeV. The effective mass $\meff$ is the sum of the 
$p_T$ of the leading three jets, the $p_T$ of the lepton and $\met$. The cut on 
$\met $ thus scales with the total transverse momentum in the event. 
The smallest of the
azimuthal separations between the jets and missing transverse energy vectors $\ourdeltaphifull$
was required to be $>0.2$ in order to remove QCD events caused by mismeasurements or 
heavy flavour decays.
Finally, the transverse mass $M_T$, calculated with 
the lepton and the missing transverse momentum 2-vector 
\ourvecptmiss, is required to be $>100$~GeV
in order to reduce the background from W bosons.

\begin{figure}
\epsfig{figure=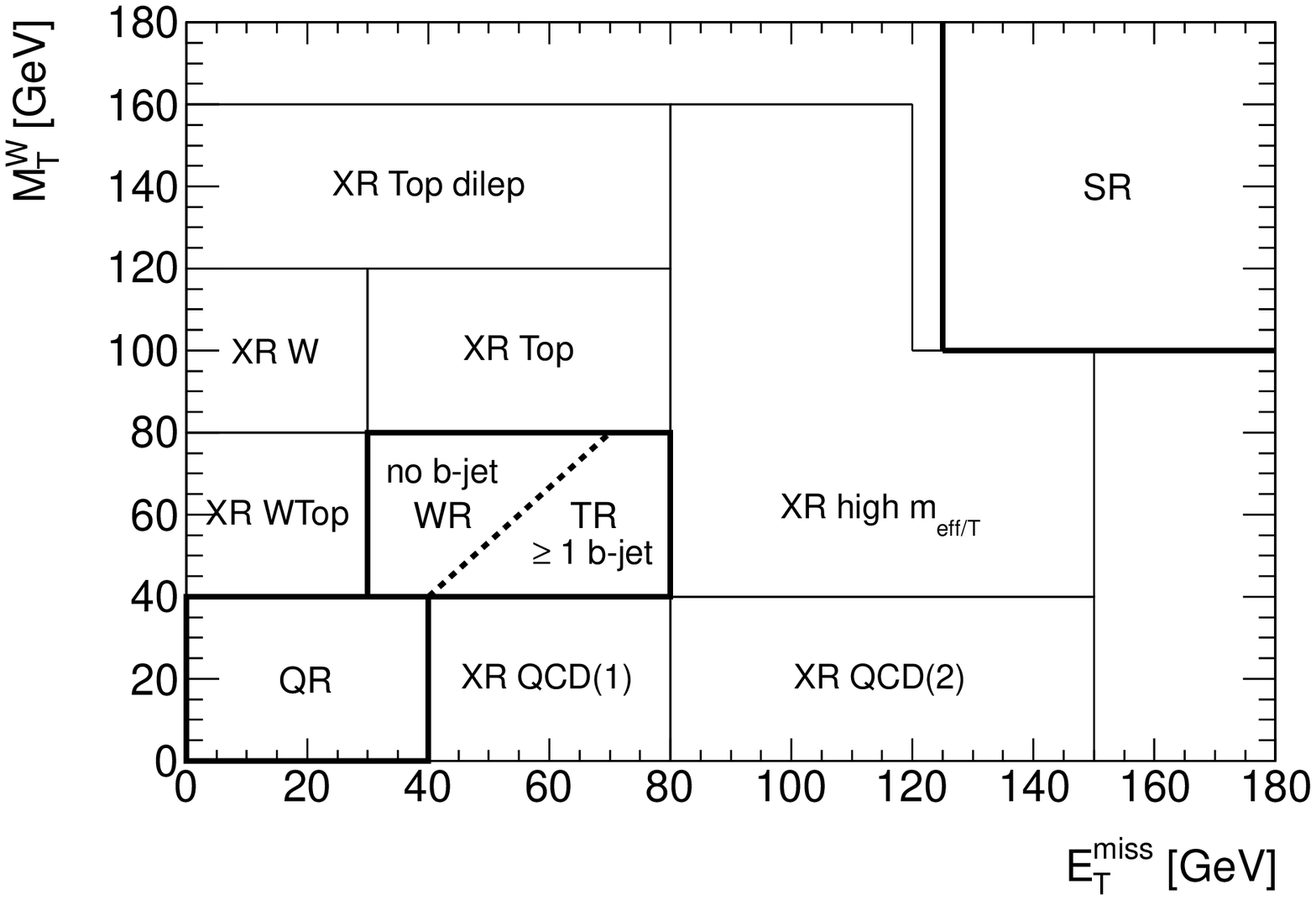,width=8cm}
\epsfig{figure=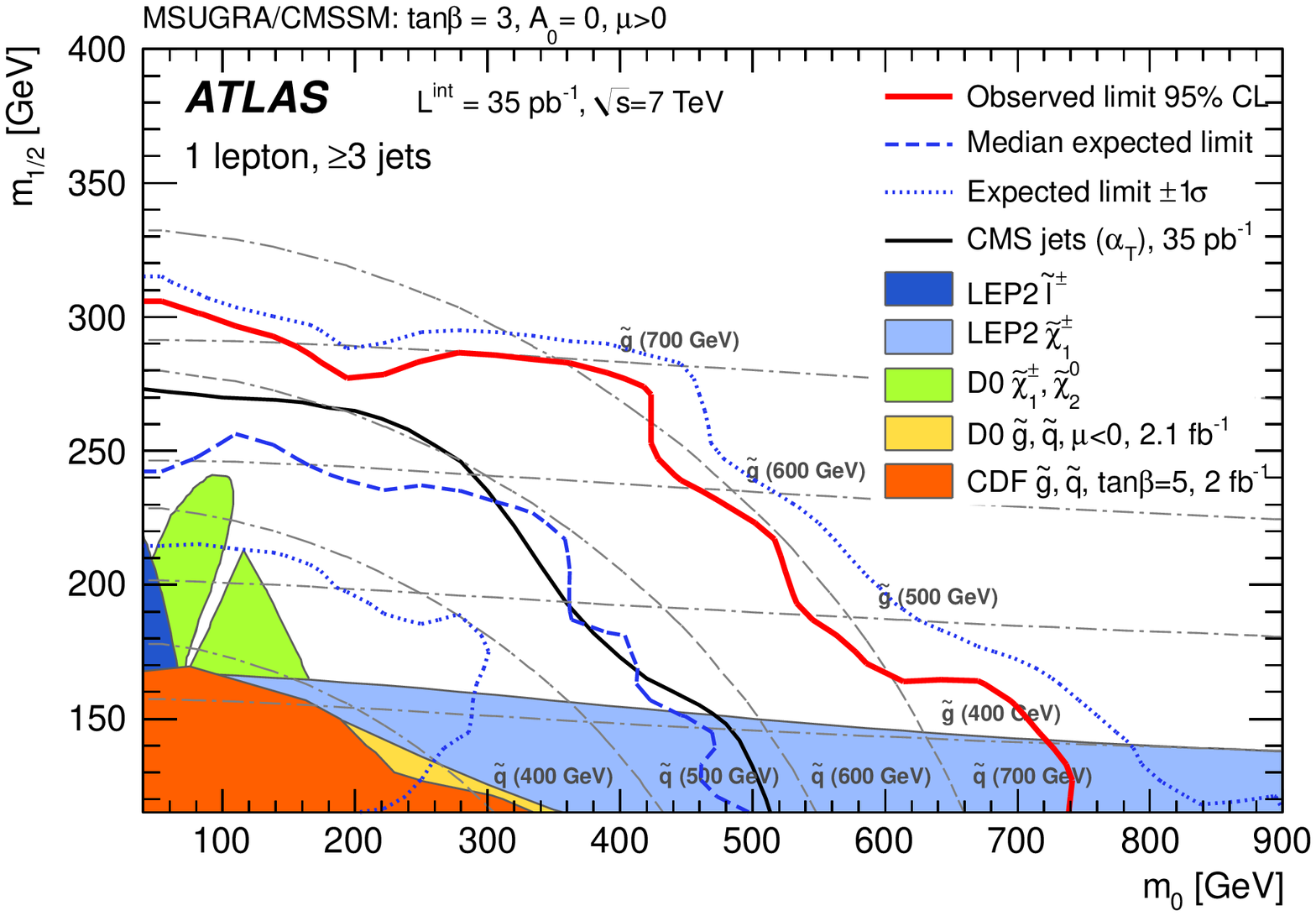,width=8cm}
\caption{The left figure shows the 
position of the signal region (SR) and the main control regions (CR) in the ($E_T^{miss}$, $M_T$) 
plane. The top enriched control region (TR) and the W+jets enriched control region (WR) are separated by the existence of a b-tagged jet candidate among the three leading jets. 
The XR regions correspond to extra validation regions. 
The right plot shows the observed and expected $95\%$ CL exclusion limits, as well as the $\pm 1 \sigma$ variation on the expected limit, in the combined electron and muon channels. 
Also shown are the published limits from CMS, CDF, D0, and the results from the LEP experiments.}
\label{fig:1lepton_limit}
\end{figure}

In order to determine the background predictions for the signal region, ATLAS has defined
several control selections. The W and top control regions e.g. are defined by keeping the 
jet and lepton selection criteria as for the signal region, but 
requiring $30<E_T^{miss}<80$~GeV and $40<M_T<80$~GeV to enhance events with W bosons.
In order to determine the amount of W+jets and top pair events separately a b-tagged jet is required
for the top control region, and a b-veto is done for the W control region.
The positions of the signal region (SR) and the main control regions (CR) in the ($E_T^{miss}$, $M_T$) plane are shown in Figure \ref{fig:1lepton_limit}. 
The transfer of the top and W normalisation factors 
measured in the control region to the signal region 
is done by Monte Carlo and the uncertainty is carefully studied, including
a validation in additional regions (see also  Figure \ref{fig:1lepton_limit}). 

In the electron selection $1.81 \pm 0.75$ events are expected and one event is found in the data.
For the muon selection $2.25 \pm 0.94$ events are expected and again only one event is found
in the data. 
The small deficit results in an observed limit which is better than the expected limit.
The limits are then derived from the profile likelihood ratio.
A model independent $95\%$ CL upper
limit on the effective cross section for new processes in the
signal region, including the effects of
experimental acceptance and efficiency, of $0.065$~pb for the
electron channel and $0.073$~pb for the muon channel is derived.
Limits are also set on the parameters of the minimal supergravity
framework, extending the limits set by the Tevatron experiment by far.  
The observed and expected upper limits are shown in Figure \ref{fig:1lepton_limit}.
For $A_0 = 0 $~GeV, $\tan \beta = 3$, $\mu>0$ and for equal squark and gluino masses,
gluino masses below $700 $~GeV are excluded at 95\% confidence level.

\section{Searches with Jets and Missing Transverse Momentum (0 lepton channel)}
The second analysis released by ATLAS only a few weeks later was the search in
channels with jets and missing transverse momentum \cite{daCosta:2011qk}.

In order to achieve a maximal reach over the
$(m_{\gluino},m_{\squark})$-plane, several signal regions are defined.
When production of squark pairs $\squark \squark$ is dominant, only
a small number of jets (one per squark from $\squark \rightarrow q \chi^0_1$) is expected. 
When production
involves gluinos, extra jets are expected from $\gluino\rightarrow q q \chi^0_1$.  
In these regions, requiring at least three jets yields better sensitivity.
For each of the four signal regions (two dijet and two three-jet selections) 
$E_T^{miss}$ is required to be $>100$~GeV.
The signal region aiming for high mass dijet events has a selection criteria
on \mttwo{}$>300$~GeV. The quantity $\mttwo$ is
a generalisation of the transverse mass for two particles
decaying to a jet and missing transverse momentum ~\cite{daCosta:2011qk}.
The other three signal regions are defined with cuts on the effective mass, which 
is here a sum over the leading two or three jets (depending on the channel) and the
missing transverse momentum.

\begin{figure}
\epsfig{figure=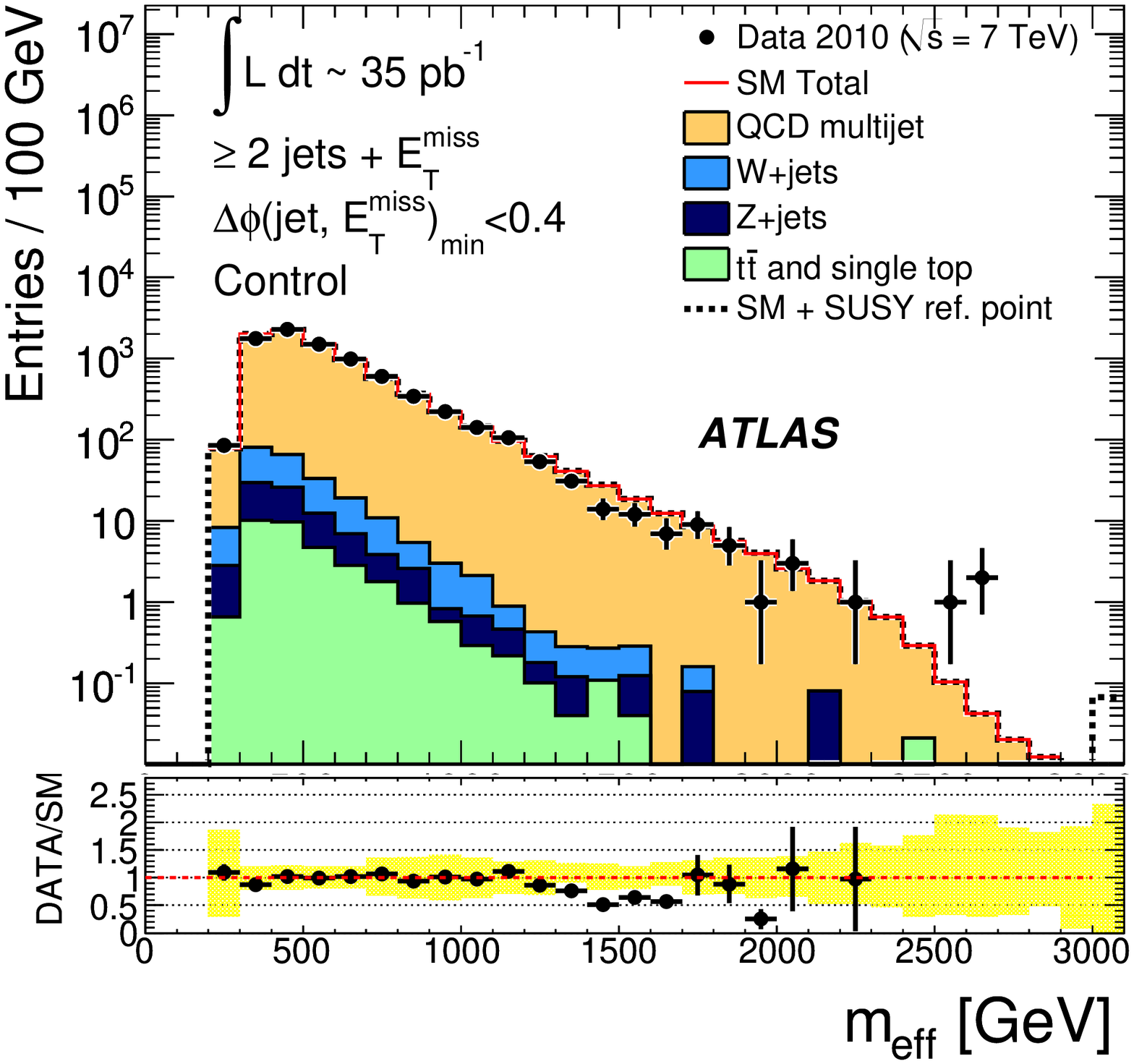,width=8cm}
\epsfig{figure=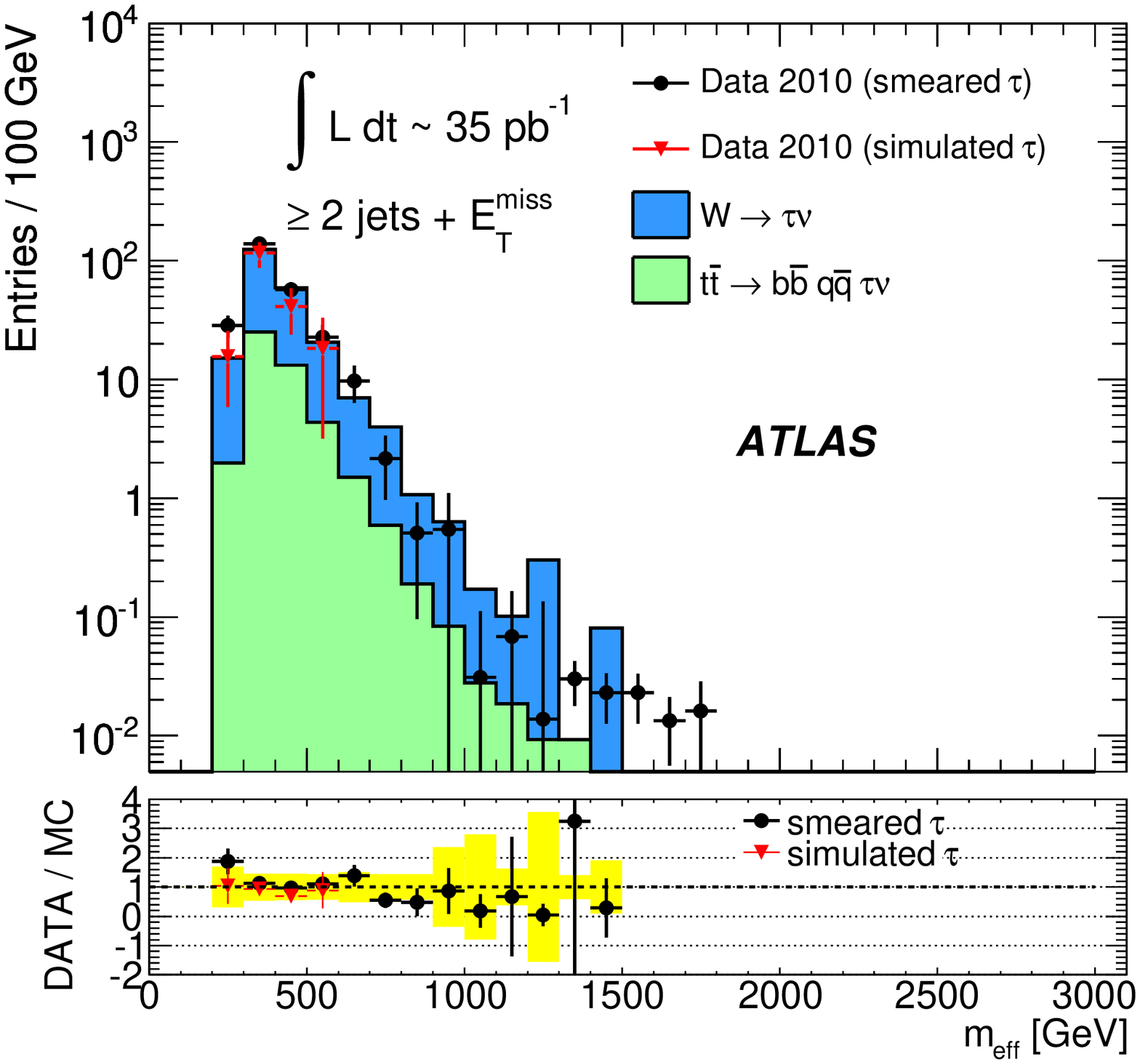,width=8cm}
\caption{
Left plot: Distribution of  $\meff$ 
after the 2-jet and missing transverse momentum 
selection in the control region defined by the reverse cut on $\ourdeltaphishort$.
Right plot: Distribution of $\meff$
for the tau-related background after the full 2-jet selection, except the cut on 
$M_{\rm{eff}}$. Two data-driven background estimates are shown, both derived 
from selected $W \rightarrow \mu \nu$ events. 
The muon is removed from the event and replaced by a Monte Carlo tau lepton decay, which is 
either smeared using resolution functions to emulate the detector response (smeared tau) or processed using the full detector simulation (simulated tau). 
}
\label{fig:0lepton}
\end{figure} 

\begin{figure}
\epsfig{figure=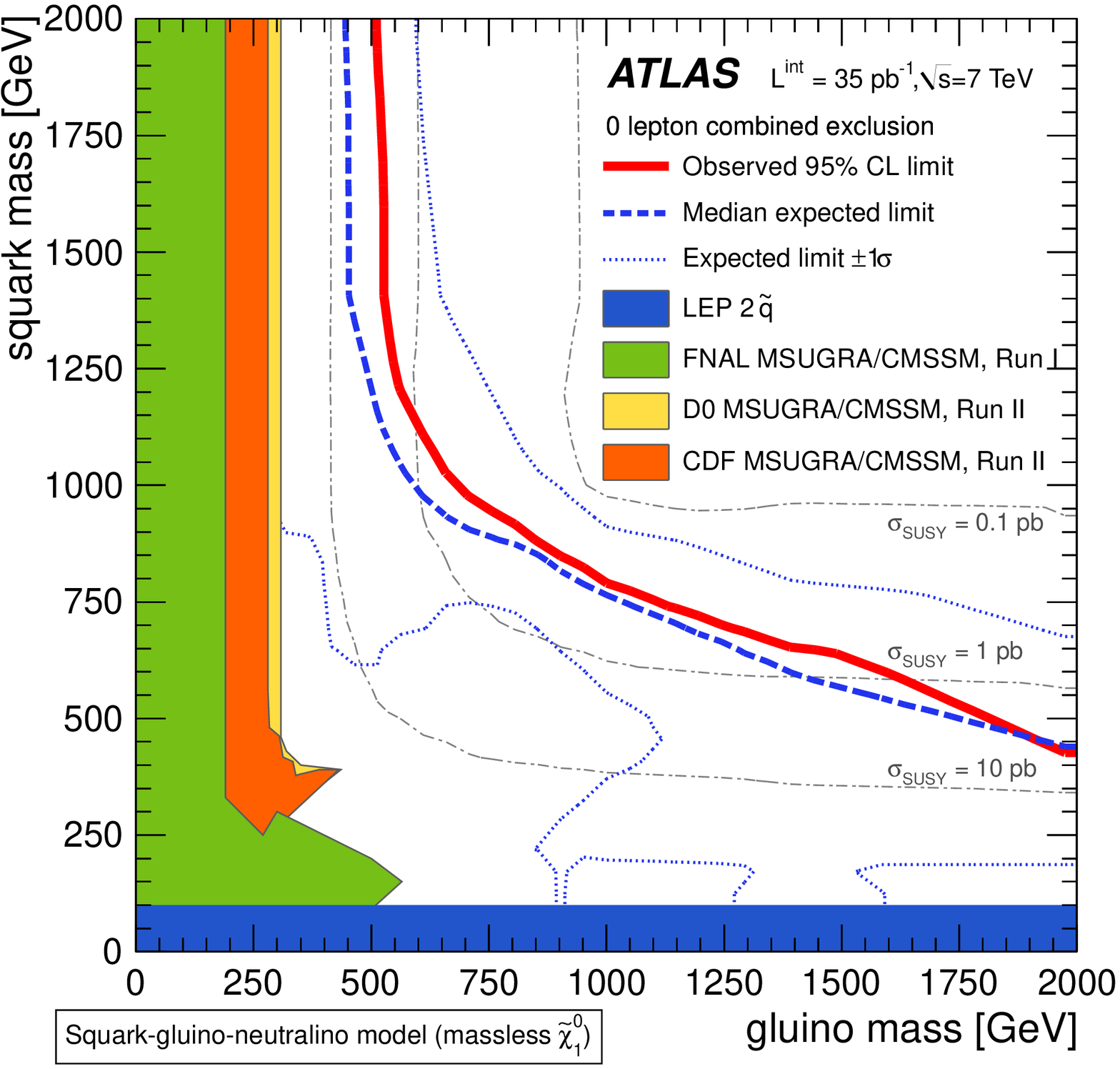,width=8cm}
\epsfig{figure=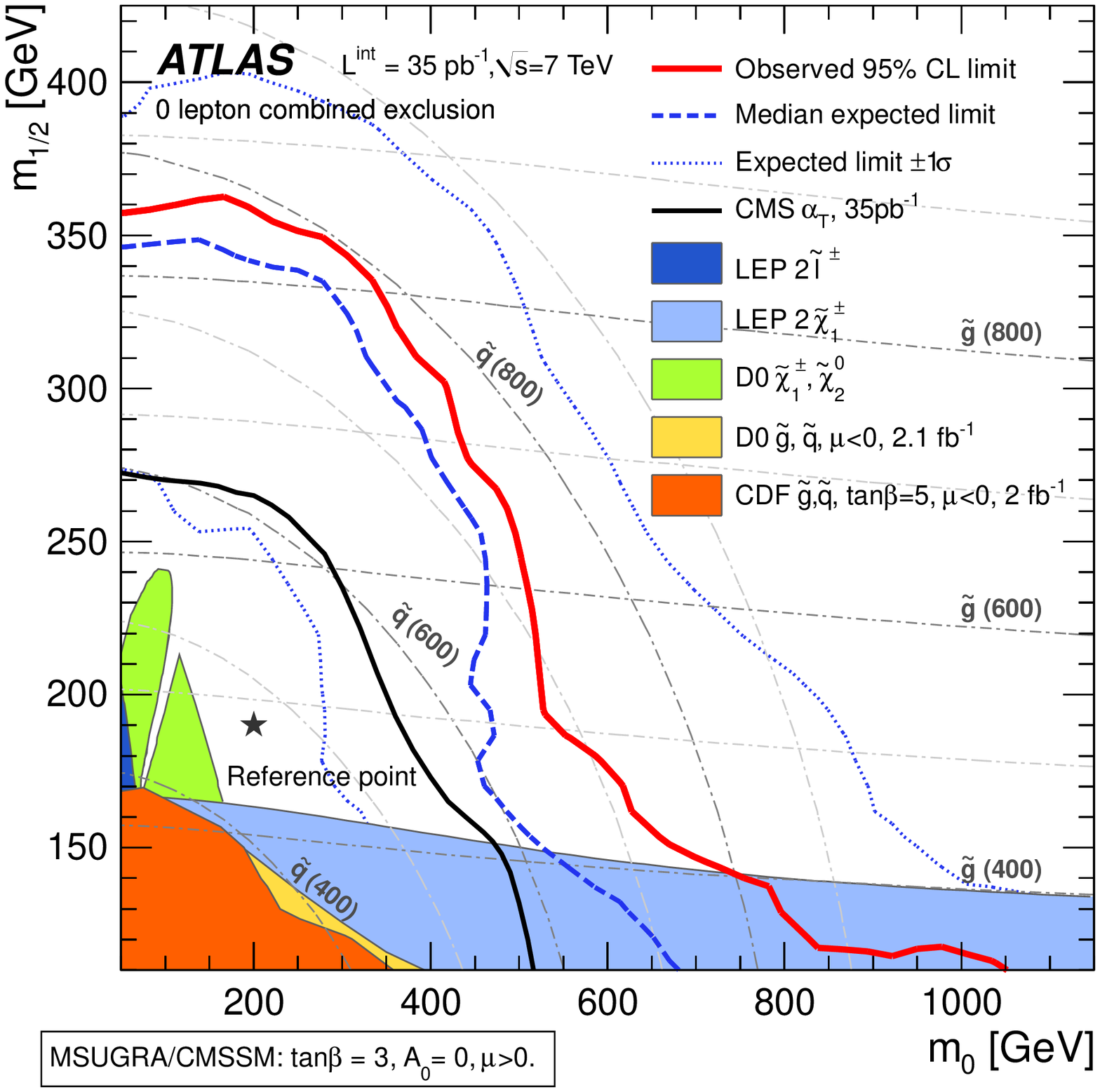,width=8cm}
\caption{
Left plot: 
$95\%$ C.L. exclusion limits in the (gluino, squark) mass plane together with existing limits for a simplified MSSM model with a massless neutralino. 
Comparison with existing limits is illustrative only as some are derived in the context of MSUGRA/CMSSM or may not assume a massless neutralino. 
Right plot:
$95\%$ C.L. exclusion limits in the $\tan(\beta)=3$, $A_0=0$ and $mu>0$ slice of MSUGRA/CMSSM, together with existing limits with the different model assumptions given in the legend. 
}
\label{fig:0leptonlimit}
\end{figure} 

The dominant SM background sources are $W+$jets, $Z+$jets, top
pair, QCD multijet and single top production.
The determination of a solid prediction for these backgrounds
in the signal regions is the main challenge in these searches.
ATLAS has carried out about $2-4$ control measurements per signal region and background.
One method to derive the QCD prediction was a 
normalisation of the QCD MC samples
by a scaling designed to achieve a match to data in control
regions obtained by reversing the $\ourdeltaphishort$ requirements.
The $\meff$ distribution in the control region is shown in Figure~\ref{fig:0lepton}.
This result was found to be consistent 
with an other data-driven estimate in which high $\ourmagptmiss$ events
were generated from data by smearing low  $\ourmagptmiss$  events on a
jet-by-jet basis with measured jet energy resolution functions. This latter technique has no MC dependencies; it provides a completely
independent determination of the QCD multijet background using only quantities measured
from the data. Additional control regions having reversed $\ourmagptmiss/\meff$ requirements were used as further checks on the normalisation.

The $Z$+jets background originates from the irreducible
component in which $Z\rightarrow\nu\bar\nu$ generates large $\ourmagptmiss$.
This background is measured in control regions were the Z decays to muons or electrons
and is also modelled with a $W$+jets control sample.
The $W$+jets background is determined via the same window in the $(M_T, \met)$ plane as in
the 1-lepton channel, but applying the jet cuts of this analysis.
Hadronic $\tau$ decays in $t \bar t \rightarrow b \bar b \tau \nu q q $
can generate large $\ourmagptmiss$ and pass the jet
and lepton requirements at a non-negligible rate.
The MC estimate for these events was checked to be
consistent with a data-driven cross-check based on
replacement of reconstructed muons in the corresponding
single lepton channels with simulated hadronic $\tau$ decays. 
The resulting $\meff$ distribution is shown in Figure~\ref{fig:0lepton}.
Agreement was also found after reweighting
the $t\bar t$ MC according to experimentally measured $b$-tag weights.

The number of observed data events and the number of SM events 
expected to enter each of the signal regions were found to be consistent in all 
four signal regions.
The signal regions exclude again non-SM effective cross sections within acceptance.
The results are interpreted in both a simplified model containing
only squarks of the first two generations, a gluino octet and a
massless neutralino, as well as in MSUGRA/CMSSM models with $\tan\beta=3$, $A_0=0$
and $\mu>0$. In the simplified model, gluino masses below $500$~GeV
are excluded at the $95\%$ confidence level with the limit increasing
to $870$ GeV for equal mass squarks and gluinos. In the MSUGRA/CMSSM
models equal mass squarks and gluinos below $775$~GeV are excluded.
Both exclusion plots are shown in Figure \ref{fig:0leptonlimit}.

\section{Searches with b-jets and missing transverse momentum}

Events with jets, one or no lepton and missing transverse momentum are 
also studied with a b-tag requirement in order to enhance the sensitivity to 
the third generation, i.e. stops and sbottom squarks \cite{Aad:2011ks}.

\begin{figure}
\epsfig{figure=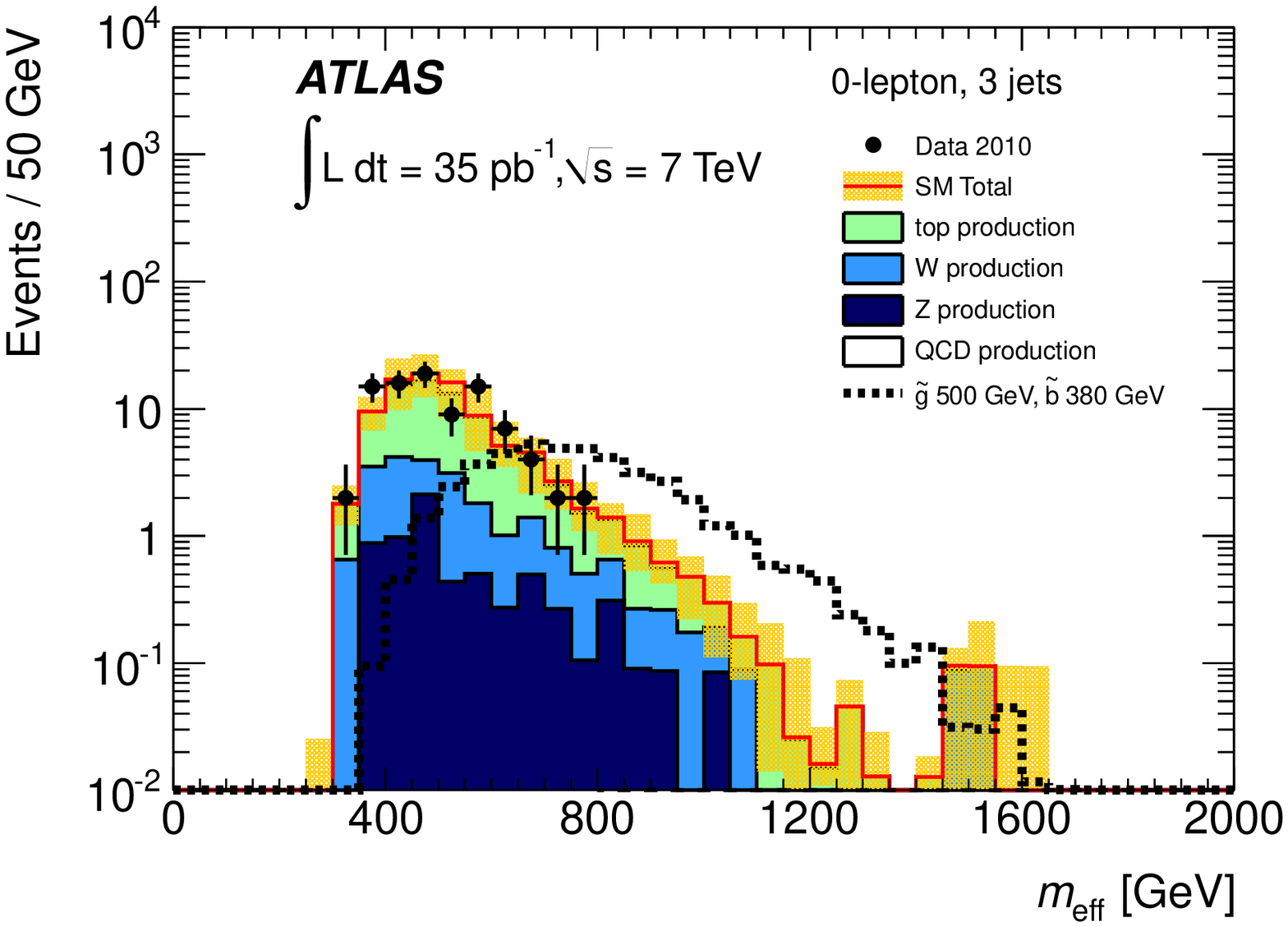,width=8cm}
\epsfig{figure=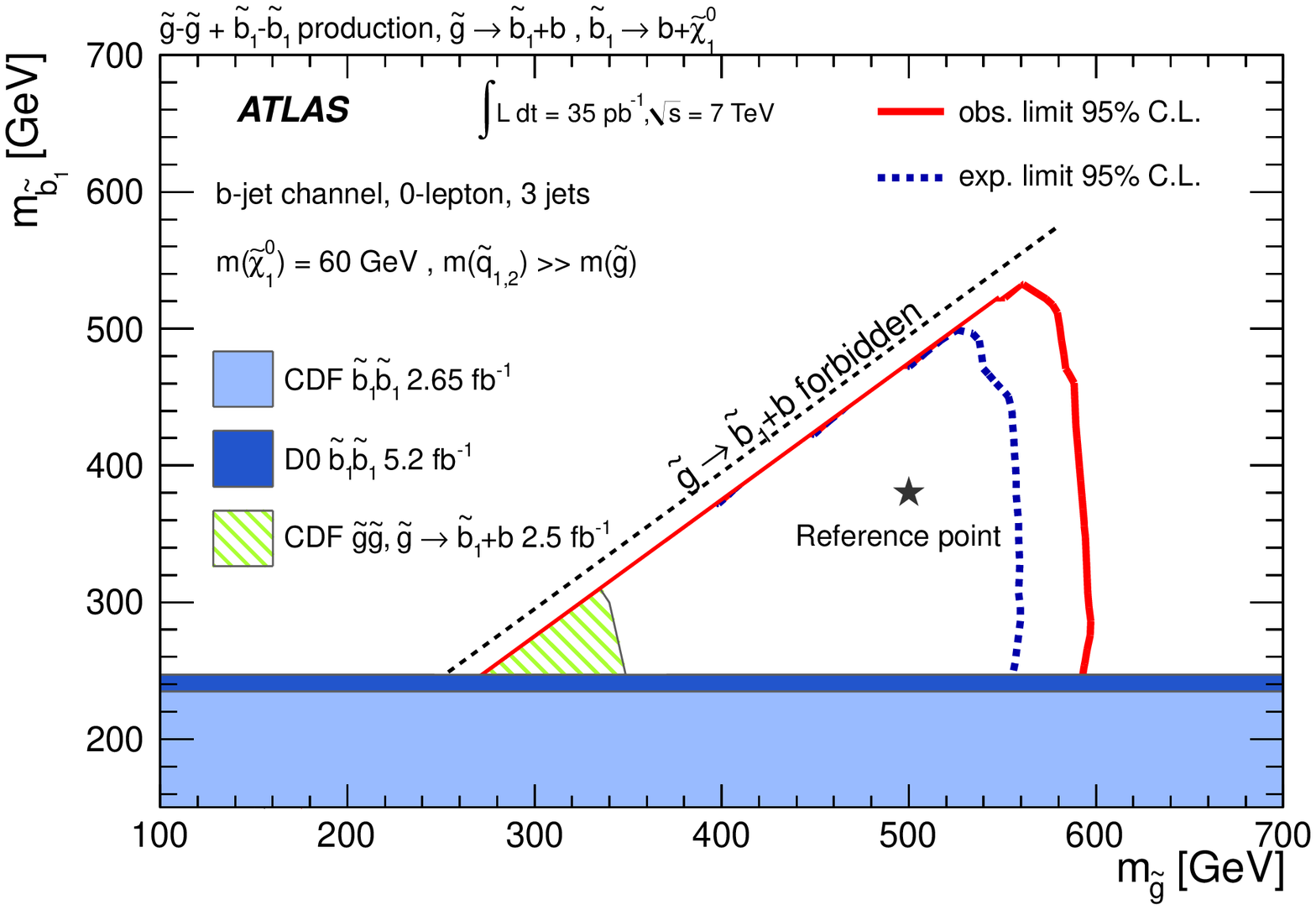,width=8cm}
\caption{
Left plot: 
Distributions of the effective mass for data and for the expectations from Standard Model processes after the baseline selections in the zero-lepton b-jet channel .
Right plot:
Observed and expected $95\%$ C.L. exclusion limits, as obtained with the zero-lepton b-jet
channel, in the ($m^{}_{\gluino},m^{}_{\sbottom1}$) plane. 
The neutralino mass is assumed to be 60~GeV. 
The result is compared to previous results from CDF and D0.
}
\label{fig:bjets}
\end{figure}

The 0-lepton b-jet selection uses slightly modified selection criteria 
compared to what was described above and is optimised 
for signals like sbottom production (either direct production or via 
gluino decay to sbottom and bottom) and the subsequent decay $\sbottom \rightarrow b \chi^0_1$.
The 1-lepton b-jet selection aims for signals of stop production
where the stop decays to a sbottom and chargino and finally the chargino 
could decay leptonically to a neutralino, a lepton and a neutrino.
The signature would therefore be b-jets, leptons and missing transverse momentum.
In both channels at least one jet is required to be b-tagged.
The dominant background source 
is then top pair production due to this extra b-jet requirement.
Again data-driven techniques have been employed to determine the QCD, W+jets and top 
backgrounds. No significant excess is observed with respect to the prediction for Standard Model processes. 
For R-parity conserving models in which sbottoms (stops) are the only squarks to appear 
in the gluino decay cascade, gluino masses below 590 GeV (520 GeV) are excluded 
at the $95\%$ C.L. The results are also interpreted in an MSUGRA/CMSSM 
supersymmetry breaking scenario with $\tan(\beta)=40$ and in an $SO(10)$ model framework.
The $\meff$ distribution for the zero-lepton selection and the resulting limit
are shown in Figure \ref{fig:bjets}.

\section{Searches with Dilepton events}
A very clean potential signal for SUSY particles could come from dilepton events.
In addition the shapes and endpoints of the dilepton mass distributions 
are a potentially perfect source to provide mass information for SUSY particles.

\begin{figure}
\epsfig{figure=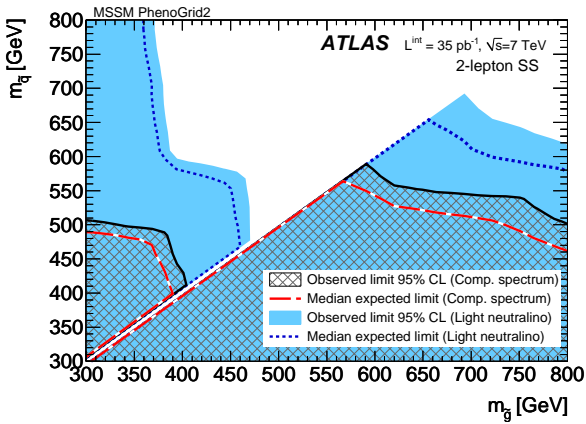,width=8cm}
\epsfig{figure=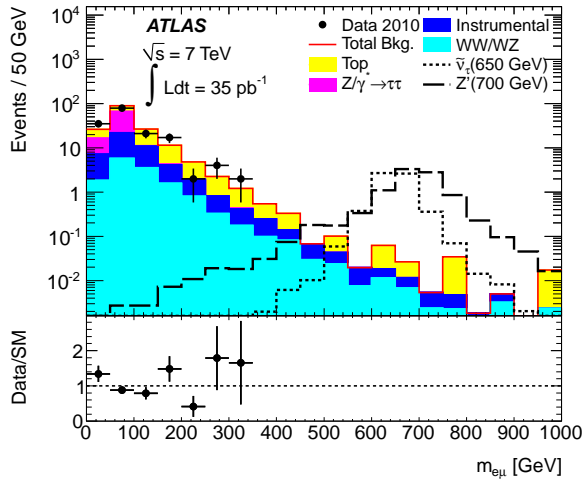,width=8cm}
\caption{
Left plot: 
Expected and observed 95\% C.L. exclusion limits in the $(m_{\tilde{g}},m_{\tilde{q}})$ plane for a MSSM models for the same sign dilepton analysis. 
Right plot:
Observed and predicted electron-muon invariant mass distributions. 
The signal simulations are shown for tau-sneutrino with a mass of 650 GeV and a Z' with a mass
of 700 GeV.
}
\label{fig:dilepton}
\end{figure}

ATLAS has searched for the production of supersymmetric 
particles decaying into final states with missing transverse momentum 
and exactly two isolated leptons.
The search strategies included events with
lepton pairs with identical sign and events with
opposite sign electric charges~\cite{Aad:2011xm}. 
The signal region for the same sign analysis is defined by $\met >100$~GeV,
the signal region for the opposite sign analysis requires $\met > 150$~GeV.
The main background of the same sign analysis arises from SM processes 
generating events containing at least one fake or non-isolated lepton.
For the opposite sign analysis the main background arises from SM top pair 
production.
The fake background is estimated by solving
linear equations to get the fake probability for a
``tight'' lepton selections via a ``loose'' lepton selection. 
Dedicated control selections are developed to estimate the background from top events.
Depending on the flavour of the two leptons ($ee,\mu\mu,\mu e$) and the 
electric charge in total 6 signal regions are used for the search. 
Again no significant excesses are observed. 
Based on specific benchmark models, limits are placed on 
the squark mass between 450 and 690 GeV for squarks approximately 
degenerate in mass with gluinos, depending on the SUSY mass hierarchy considered (see
Figure~\ref{fig:dilepton}).
The important (free) parameters of the MSSM model are the three gaugino masses and
the squark and slepton masses ~\cite{Aad:2011xm}. 
The slepton is light in order to enhance the lepton decay.

Dilepton events are also studied with the so called ``flavour subtraction'' method
~\cite{Collaboration:2011xk}. 
Flavour uncorrelated backgrounds are 
subtracted using a sample of opposite flavour lepton pair events. 
The dominant background from top pair production can be subtracted via this method.
In SUSY events the production of the two leptons can be correlated, 
if lepton flavor is conserved in the decays of e.g. a heavy neutralino
to a slepton and a lepton and subsequently to $ll \chi^0_1$.
The search for 
high missing transverse momentum events containing opposite charge identical 
flavour lepton pairs yields no significant
excess. Here limits are set on 
the model-independent quantity $S$ , which measures the
mean excess from new physics taking into account 
flavour-dependent acceptances and effiencies.

A third generic 
search is performed for heavy particles decaying into an electron-muon final state ~\cite{Aad:2011kt}.
Again the fake backgrounds are carefully determined via data-driven techniques.
No excess above the Standard Model background expectation is observed (see Figure~\ref{fig:dilepton}). 
Exclusions at $95\%$ confidence level are placed on two representative models. 
In an R-parity violating supersymmetric model, tau sneutrinos 
with a mass below $0.75$ TeV are excluded, assuming 
single coupling dominance and the R-parity violating 
couplings to be $\lambda'_{311}=0.11$, $\lambda_{312}=0.07$ (in order to compare with
previous Tevatron studies).
The ATLAS results extend to higher mass then previously studied at the Tevatron.

\section{Searches for new slow-moving massive particles}
ATLAS has also pursued a
search for slow-moving charged particles (SMPs)~\cite{Aad:2011yf}. 
Such particles are expected in some new physics models where the 
new (SUSY) particles are not allowed to decay, e.g. because 
the decay goes via highly virtual particles or because the couplings are very small.
If these long lived particles are squarks or gluinos they will 
hadronize and form so called R-hadrons. The signal of such heavy particles
will be a slowly moving heavy hadron.

The ATLAS detector contains 
a number of subsystems which provide information which can be used
to distinguish SMPs from particles moving at velocities close to the speed of light. 
Two complementary
subsystems used in this work are the pixel detector, which measures ionisation energy loss (dE/dx),
and the tile calorimeter, which measures the 
time-of-fight from the interaction point for particles which
traverse it. 

The events are triggered by a $\met$ and track requirement. 
For each event, the mass is estimated by dividing its momentum by $\beta \gamma$ , 
determined either from
pixel detector ionisation or from the tile calorimeter.
Rather than relying on simulations to predict the tails of the Pixel and Tile beta 
distributions, a data-driven method is used to estimate the background.
Estimates for the background distributions are
obtained by combining random momentum values 
(after the kinematic cuts) with random
measurements of the Pixel and Tile $\beta$. This works since no correlation was observed
in these three measurements.
                                                                                 
\begin{figure}
\epsfig{figure=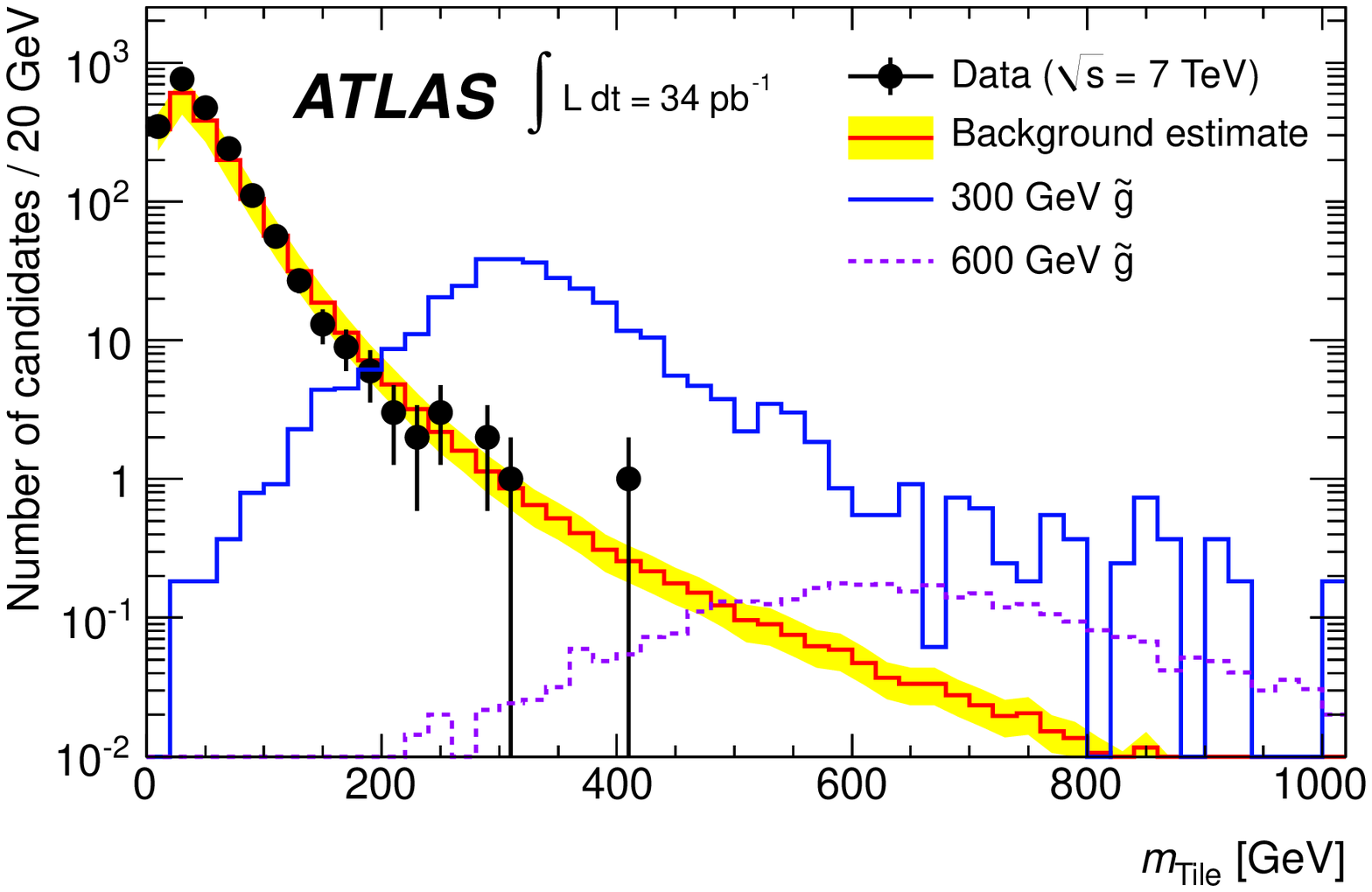,width=8cm}
\epsfig{figure=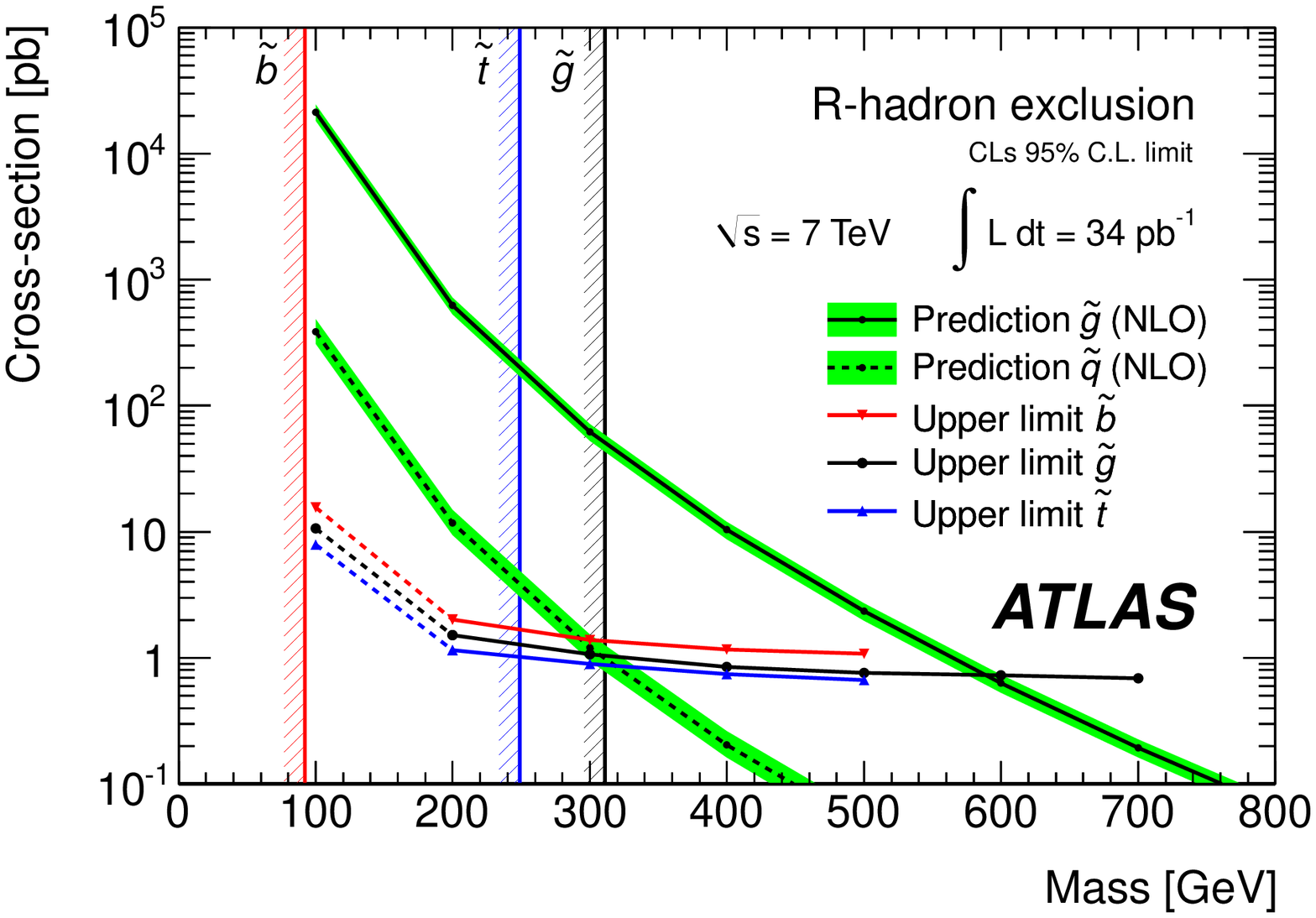,width=8cm}
\caption{
Left plot: The reconstructed mass of the R-hadron candidate and the 
background estimates for the tile calorimeter. 
Signal samples are superimposed on the background estimate. 
Right plot:
Cross-section limits at $95\%$ CL as a function of sparticle mass. 
The mass limits quoted in the text are inferred 
by comparing the cross-section limits with the model predictions. 
Previous mass limits are indicated by shaded vertical lines for 
sbottom (ALEPH), stop (CDF) and gluino (CMS). 
}
\label{fig:rhadron}
\end{figure}

Using data combined from these independent measurements, 
there are no events containing a candidate with mass greater than 100 GeV.
This result is interpreted in a framework of supersymmetry models with R-hadrons and  
$95\%$ CL limits are set on the production cross-sections of squarks and gluinos. 
The influence of R-hadron interactions in matter was studied using a number of different models, 
and lower mass limits for stable sbottoms and stops are found to be 
294 and 309 GeV respectively (see Figure \ref{fig:rhadron}).  
The lower mass limit for a stable gluino lies in the range from 
562 to 586 GeV depending on the model assumed.
Each of these constraints is the most stringent to date.

\section{Summary and Conclusion}
The ATLAS experiment has started to search for signals of Supersymmetry. 
Due to the large cross sections for squarks and gluinos the sensitivity of the
2010 LHC data exceeds by far that of all previous collider experiments.
The search is performed in a variety of different final states. 
ATLAS has presented the most stringent limits to date in many scenarios and
no signal has been observed yet.
The next years will likely allow a conclusive answer about 
the fate of low mass Supersymmetry.

\section*{Acknowledgments}
The author acknowledges the support by the Landesstiftung Baden W\"urttemberg and the BMBF.


\section*{References}
\begin{thebibliography}{99}

\bibitem{Martin:1997ns}
  S.~P.~Martin,
  arXiv:hep-ph/9709356.

\bibitem{Aad:2008zzm}
  ATLAS Collaboration,
  JINST {\bf 3} (2008) S08003.


\bibitem{Aad:2011hh}
  ATLAS Collaboration,
  PRL 106, 131802 (2011), arXiv:1102.2357 [hep-ex].


\bibitem{daCosta:2011qk}
  ATLAS Collaboration,
  accepted by PLB, arXiv:1102.5290 [hep-ex].

\bibitem{Aad:2011ks}
  ATLAS Collaboration,
  submitted to PLB, arXiv:1103.4344 [hep-ex].


\bibitem{Aad:2011xm}
  ATLAS Collaboration,
  submitted to EPJC letters, arXiv:1103.6214 [hep-ex].

\bibitem{Collaboration:2011xk}
  ATLAS Collaboration,
  accepted by EPJC letters, arXiv:1103.6208 [hep-ex].

\bibitem{Aad:2011kt}
  ATLAS Collaboration,
  accepted by PRL, arXiv:1103.5559 [hep-ex].




\bibitem{Aad:2011yf}
  ATLAS Collaboration,
  accepted by PLB, arXiv:1103.1984 [hep-ex].



\end{thebibliography}
\end{document}